\documentstyle[12pt]{article}
\begin{document}

\title{ \begin{flushright}
{\normalsize UMDEPP 00-47}\\
{\normalsize VPI-IPPAP-99-12}\\ \ \\
\end{flushright}
Searching for Supersymmetry in Hadrons}
\author{ S. James Gates, Jr.
\\
\\
     \it{Department of Physics}\\
     \it{University of Maryland}\\
     \it{College Park, Maryland 20742-4111}\\
     \\
     and\\
     \\
     Oleg Lebedev
     \\
     \it{Department of Physics}\\
     \it{Virginia Polytechnic Institute and State University}\\
     \it{Blacksburg, Virginia 24061-0435}\\ \ \\}
\maketitle
\thispagestyle{empty}

\begin{abstract}
We discuss the possibility of the existence of a long-lived top squark
($\tau\ll 10^{10}\;years$) and its motivation.  If the stop is indeed
metastable, it forms hadrons. We study properties of the low-energy
stop-containing hadrons and their signatures in collider experiments.\\
PACS: 12.60.J, 14.80.Ly, 14.20
\end{abstract}

\newpage

\section{Introduction}

${~~~}$ A feature of most discussions on the possibility of
observing supersymmetry in hadronic physics is that attention
has been paid almost exclusively to ``fragmentation physics.''
That is predictions have been made of what distinctive signature
might occur in hadronic jets due to the production and subsequent
decay of sparticles.  This appears due to the expectation that
the mass of the lightest squarks will be substantially greater
than that of the lightest super-partners of standard model
particles that do not carry color.  This is the pattern within
the conventional standard model and is so expected for its
presumptive supersymmetric successor.  Under this circumstance,
bound states  involving squarks and ordinary quarks would not
undergo ``hadronization,'' superpartner hadrons (``$shadrons$'')
would never form and hadron spectroscopy would be irrelevant.

Although unconventional, we can ask a simple question,  ``If the mass
of squarks were comparable to that of the LSP (lightest supersymmetric
particle), what type of hadrons might be expected?''  Part of what
prompts us to ask this unconventional question has been our formal
investigations of possible structures that give naive supersymmetric
extensions \cite{pionini} of the effective field theory describing
low-energy pion physics.  These results seem to indicate that
supersymmetric extensions of these are formally possible.

The possibility of the existence of supersymmetric hadrons was first
pointed out by Farrar and Fayet \cite{fayet}.
Historically, models containing light gluinos rather than squarks
appeared
to be better motivated.
Subsequently, the gluino containing hadrons (``$R$-$hadrons$'')
have been studied to a great extent \cite{farrar}. On the other hand,
another option - the squark containing hadrons - has
not drawn similar attention. We believe that, in the
ongoing search for supersymmetry, one cannot afford the luxury
of ignoring this possibility however unlikely it may seem. In this
letter, we attempt, at least partially, to fill this gap
in the phenomenology of supersymmetry.

\section{Metastable Stop}
${~~~}$ In this paper we study the possibility of a metastable top
squark.
It is well known that the top squark can be the lightest supersymmetric
particle (LSP) if  the left-right stop mixing is comparable to the
squark masses (for reviews, see \cite{kane}). Combined with the
requirement that the interactions be strictly R-parity invariant,
this results in the stability of the top squark as the stop decays
into other superparticles are not allowed kinematically.

It is, however,
also well known that such a scenario encounters serious cosmological
problems. A stable stop would form stable hadrons. The abundance of
such hadrons can be estimated in the standard cosmological model knowing
the masses and cross sections of these particles. One finds that stable
100 $GeV$ hadrons  should exist in concentrations of one in
$10^{10}-10^{
12}$ nucleons \cite{wolfram}. Thus they could be detected as anomalously
heavy protons or, if they get bound in nuclei, they could be found in
anomalous isotopes \cite{dover}. However, years of experimental searches
for anomalous protons and isotopes have resulted in extremely tight
bounds
on their concentration in matter \cite{smith},\cite{hemmick}:
\begin{eqnarray}
&&n_{q=+1}/n_{nucleon} <  10^{-29}\;,\nonumber\\
&&n_{q=-1}/n_{nucleon} < 4\times 10^{-20}\;
\end{eqnarray}
for m=100 $GeV$; the displayed bound on $n_{q=-1}$ results from the
search
for anomalous isotopes of carbon.  Apparently, these numbers are in
conflict with the expected concentration of the stop containing hadrons.
Thus, the top squark cannot be absolutely stable.

The stability of the stop arises from the assumption of an exact
R-invariance.  However, recently, with the discovery of the neutrino
mass,
this  assumption has lost some of its theoretical motivation.  It seems
more plausible that R-parity is not an exact symmetry as is not 
lepton (baryon) number (see \cite{dreiner} for a review). In this case
the LSP decays into ordinary particles. If R-parity is broken only
slightly,
the decay time can be very long. This may lead to other cosmological
problems. For example, if the decay products contain quarks, the balance
between protons and neutrons would be broken during the nucleosynthesis
leading to unacceptable abundances of helium and other light elements
\cite{reno}. This places strict constraints on the lifetime of the dark
matter candidates such as neutralinos and gravitinos. These
considerations
can also lead to significant constraints on the lifetimes of new
strongly
interacting particles if one assumes that the baryon asymmetry in the
``new'' sector is the same as it is in the regular matter sector.
This assumption, however, is strongly model-dependent and needs not be
the case. In particular, if the asymmetry in the stop sector is much
smaller than that in the matter sector, the concentration of the susy
hadrons at the time of nucleosynthesis is too small ($\sim
10^{-10}n_{bar}$ \cite{wolfram}) to affect the formation of light
elements. In this case the stop lifetime is left essentially
unconstrained
as long as
\begin{eqnarray}
&& \tau_{stop} \ll 10^{10}\;years\;.
\label{eq:life}
\end{eqnarray}
The top squark with such a lifetime is the subject of our paper.

The stop is allowed to decay into quarks and charged leptons
via the R-breaking interactions:
\begin{equation}
W_{\not{R}}
=  \lambda_{ijk}'  \hat{L}_i \hat{Q}_j \hat{D}_k
+  \lambda_{ijk}'' \hat{U}_i \hat{D}_j \hat{D}_k\;,
\label{eq:superpotential}
\end{equation}
where $\hat{L}_i$, $\hat{Q}_i$, $\hat{U}_i$, and
$\hat{D}_i$ are the usual MSSM superfields. The corresponding decay
width
is given by
\begin{equation}
\Gamma = {\vert \lambda'\;,\;\lambda''\vert^2\over 16\pi}m_{\tilde
t}\;.
\end{equation}
The requirement that all the stops have decayed by now can then be
translated into the lower bound on the couplings:
\begin{equation}
\lambda',\lambda'' \gg 10^{-21}\;
\label{eq:bound}
\end{equation}
with $m_{\tilde t}=100\;GeV$.
The most recent upper bounds on these couplings from particle
experiments
can be found in \cite{allanach}.

A model containing such small numbers may seem ill-motivated. However,
very small R-parity violation can naturally arise in some flavor models
with a discrete symmetry \cite{hall},\cite{carone}.  In particular,
in the model of  Ref.\cite{carone}, the $(S_3)^3$ symmetry prohibits
renormalizable R-breaking operators. R parity arises as an accidental
symmetry and can be broken by nonrenormalizable operators induced at
the Plank scale. Then the effective low-energy trilinear operators get
suppressed by $(M_f/M_{Pl})^2$ and $M_f/M_{Pl}$, where $M_f$ is the
flavor
symmetry breaking scale. With a sufficiently low $M_f$, the R-breaking
couplings in the range given by Eq.\ref{eq:bound} can be naturally
produced.
For example, the lower bound on the $\lambda',\lambda''$ corresponds to
$M_f\sim 10^9\; GeV$ if the R-breaking interactions are suppressed by
$M_{Pl}^2$. This provides a theoretical motivation for studying a
possibility of a long-lived top squark.

\section{Lightest Stop Containing Hadrons}

${~~~}$ Since any observable object has to be a color singlet, a
metastable top squark must be confined in hadrons. Within
${\cal{O}}(10^{-23})sec$ it will form a bound
state with either an antiquark - $mesonino$, or a couple of quarks -
$sbaryon$. Clearly, the lightest mesonino (sbaryon) is just as stable
as the stop itself. Masses of the lightest $shadrons$ can be
approximated
within $1\;GeV$ by the stop mass which we will set to $100\;GeV$,
however
splittings between them will be very important for phenomenology.
The natural candidates for the lightest hadron are listed in Table 1
(this is to be supplemented with the charge conjugate states).

\begin{table}
\begin{center}
\begin{tabular}{|c||c|c|c|c|c|}
\hline
& quark-squark content & spin & charge & isospin & baryon number \\
\hline
$T^0$ &$\tilde t \bar u$ & 1/2 & 0 & 1/2 & 0 \\
$T^{+}$ &$\tilde t \bar d$ & 1/2 & +1 & 1/2 & 0 \\
$V^{0}$ &$\tilde t dd$ & 1 & 0 & 1 & 1\\
$V^{+}$ &$ {1\over\sqrt{2}}\tilde t (ud+du)$ & 1 & +1 & 1 & 1\\
$V^{++}$ &$\tilde t uu$ & 1 & +2 & 1 & 1\\
$S^{+}$ &$ {1\over\sqrt{2}}\tilde t (ud-du)$ & 0 & +1 & 0 & 1\\
\hline
\end{tabular}
\end{center}
\caption{Lightest hadrons containing a top squark.}
\end{table}

In the presence of isospin symmetry breaking, $T^0$ is not a mass
eigenstate.  Indeed, due to the diagram in Fig.1, $T^0$ mixes with
$\bar T^0$ forming  Majorana fermions $T_1$ and $T_2$.

Let us estimate how large is the mixing.  The leading contribution is
given by the gluino diagram in Fig.1. Further, in the limit $m_{\tilde
g}^2
\gg m_{\tilde t}^2$, the mixing is dominated by the diagram with a
gluino
mass insertion (Fig.2). It is also reasonable to assume that the
lightest
stop is predominantly right handed. Then, the generated effective
Hamiltonian
is given by
\begin{equation}
{\cal{H}}_{eff}\approx {g_s^2 \tilde V_{tu}^2 \over 3 m_{\tilde g}}
(\bar u_R \tilde t_R)^2 \;,
\end{equation}
where $\tilde V$ is the mixing matrix at the gluino vertex. To convert
this interaction into a mass term, one needs to know the matrix element
$\langle T^0\vert {\cal{H}}_{eff} \vert \bar T^0 \rangle$. In the case
of the Standard Model $K-\bar K$ mixing, the needed matrix elements
can be derived from the PCAC relation and the vacuum saturation
assumption. In our case we will simply parametrize the matrix element
assuming
\begin{equation}
\langle T^0\vert (\bar u_R \tilde t_R)^2 \vert \bar T^0 \rangle=F^2,
\end{equation}
where $F$ has dimension of mass{\footnote{In our convention
$\langle T^0\vert  T^0 \rangle= \delta^{(3)}(p-p')$.}}.
The size of $F$ can be estimated by the characteristic hadronic matrix
element of a heavy quark meson $(f_B \;m_B^2)^{1/3}\sim
{\cal{O}}(1)\;GeV$
since this combination is constant in the heavy quark limit.

As a result, we get the following mass matrix in the $(T^0,\bar T^0)$
basis:
\begin{eqnarray}
&&M= \pmatrix{m&\epsilon \cr
              \epsilon^{*} & m \cr}\;,
\end{eqnarray}
where $\epsilon = {g_s^2 \tilde V_{tu}^2 \over 3 m_{\tilde g}} F^2$.
In principle, $\tilde V_{tu}$ may contain a complex phase, leading to
CP violation in mixing. However, in this paper, we restrict ourselves to
the CP conserving case. The consequent mass eigenstates are maximal
mixtures
of $T^0$ and $\bar T^0$:
\begin{eqnarray}
&&T_1 = {1\over\sqrt{2}}(T^0-\bar T^0)\;,\;m_1=m-\epsilon\;,\nonumber\\
&&T_2 = {1\over\sqrt{2}}(T^0+\bar T^0)\;,\;m_2=m+\epsilon\;.
\end{eqnarray}
It is interesting to compare the mass splitting between the two
eigenstates to a typical mass splitting in the neutral meson systems.
For
example, in the Standard Model, the $B^0-\bar B^0$ mixing arises only
through
a box diagram resulting in  $(\Delta m/m)_{B^0} \sim 10^{-13}$. In
contrast,
the $T^0-\bar T^0$ mixing exists already at the tree level and
\begin{equation}
{\Delta m\over m}= {2\over 3} g_s^2 \tilde V_{tu}^2 {F^2\over
m_{\tilde g} m_{\tilde t}} \sim 10^{-9}
\end{equation}
for $\tilde V_{tu} \sim V_{31}^{CKM},\;m_{\tilde g}\sim 300\;GeV,
\; m_{\tilde t}\sim \;100 GeV $ and $F\sim 1\; GeV$. The result, of
course,
is sensitive to $F$, but it is clear that $\Delta m/m$ is enhanced by
orders of magnitude over that in the Standard Model.

Let us now turn to the discussion of the isospin partner of $T^0$ -
$T^+$.
Since the top squark can be treated as a spectator, the mass splitting
in this isospin doublet can be attributed to the $m_d-m_u$ mass
difference and the corresponding Coulomb interactions.
As a result, $T^+$ is unstable and $\beta$-decays:
\begin{equation}
T^+ \;\rightarrow\; T^0\; +\; e^+ \nu
\end{equation}
(It is understood that $T^+$ actually decays into a linear combination
of the mass eigenstates $T_1$ and $T_2$.) Clearly, this process is
governed by the usual Fermi interaction in analogy with a neutron
$\beta$-decay:
\begin{equation}
H = {G\over \sqrt{2}}\; {\overline T^+}\gamma^{\mu} (1-\alpha
\gamma_5)T^0\;
{\overline e}\gamma_{\mu} (1-\gamma_5)\nu\;,
\end{equation}
where $\alpha \sim 1$ accounts for hadronic effects. The decay width is
given by that of a neutron $\beta$-decay:
\begin{equation}
\Gamma \sim {G^2 \Delta^5 \over 60 \pi^3}
\end{equation}
with $\Delta = m_{T^+} - m_{T^0} \sim m_{D^+} - m_{D^0} \approx 5 MeV.$
The corresponding decay time is about 3 $sec$. So within a few seconds
all of the $T^+$'s will decay into the $T^0$'s{\footnote{This estimate
is
sensitive to the exact value of the mass difference. The actual
decay \newline ${~~~~~}$ time may be an order of magnitude larger. See
\cite{cahn} for similar considerations.}}.

One can go further in eliminating candidates for the lightest shadron.
Let us consider the mass splitting between the vector isotriplet $V$ and
the scalar isosinglet $S^+$. The mass difference arises due to the spin
flip of one of the quarks. This effect is accounted for in the quark
potential model via the spin-spin interaction (see, for example,
\cite{don}).
Such an interaction is part of the one gluon exchange potential and is
responsible for the $\rho-\pi$ and $\Delta-N$ mass differences.
Again, treating a spin 0 stop as a spectator, we can write the
hyperfine interaction as follows:
\begin{equation}
H_{h.f.}= k_{qq} \;\vec{s_1} \cdot \vec{s_2}\;\delta^{(3)}(\vec{r})\;.
\label{eq:hyp}
\end{equation}
The constant $k$ depends on the quark masses and on whether two quarks
or
a quark and antiquark are interacting, $k_{qq}= {1\over2} k_{q\bar q}$.
The scalar product $\vec{s_1} \cdot \vec{s_2}$ is $1/4$ for the vector
state and $-3/4$ for the scalar state. This situation is very similar to
what we encounter in the $\rho-\pi$ system up to the substitution
$k_{qq}\rightarrow k_{q\bar q}$. Therefore,
\begin{equation}
m_V - m_{S^+} \sim {1\over2} (m_{\rho} - m_{\pi}) \approx 300\; MeV\;.
\end{equation}

Another way to estimate the $V-S^+$ mass difference is to exploit the
heavy quark symmetry of the strong interactions \cite{isgur}. In the
heavy quark limit, the interactions of a heavy quark are independent
of its spin and flavor. Thus the heavy quark spectroscopy depends on
the light quark configuration only. Using this fact, one can approximate
the $V-S^+$ mass difference by that of the $\Sigma$ and $\Lambda$
baryons containing a heavy quark. This gives
\begin{equation}
m_V - m_{S^+} \sim 200\; MeV\;,
\end{equation}
which is about 30 \% away from the previous estimate{\footnote{To obtain
this number, we use the mass of $\Lambda_c$ and the average mass of
$\Sigma_c$, spin 1/2  \newline ${~~~~~}$ and $\Sigma_c$, spin 3/2 (which
should be the same in the heavy quark limit) \cite{pdg}. Anologous 
\newline ${~~~~~}$ information for the $b$-baryons currently is not
available.}}.

In both cases the mass splitting is well above the pion threshold,
so $V$ will decay into $S^+$
within about $10^{-23}$ $sec$:
\begin{equation}
V\;\rightarrow\; S^+\; +\; \pi
\end{equation}

Thus we are left with only two potentially long-lived shadrons - $T^0$
and
$S^+$.  In principle, there is a possibility that one of them will decay
into another. Since they carry different baryon numbers, a proton must
be
emitted in such a decay: $S^+\rightarrow T^0+p$. It is, however, quite
unlikely that the $S^+-T^0$ mass difference is above $1\;GeV$.
To see this, we can again exploit the spin independence of the heavy
quark
interactions and use an analogy with the heavy quark hadrons.
The $S^+-T^0$  mass splitting can be compared to that for
the  $\Lambda_{b(c)}$ baryons and  $B(D)$ mesons.
This estimate yields
\begin{equation}
m_{S^+} - m_{T^0} \sim 400\; MeV\;.
\end{equation}
Thus, the decay $S^+\rightarrow T^0+p$ is not allowed kinematically.

To summarize, we have argued that there are four long-lived shadrons
(including the charge conjugate states):  $T_{1,2}$ and $S^{\pm}$. Their
lifetimes are given by the stop lifetime. Other light shadrons decay
into
these states either strongly or via $\beta$-decay.

\section{Possibility of Detection}

${~~~}$ Supersymmetric particles have been the subject of collider
searches for many years (for recent reviews see \cite{carena} and
references therein). In most of these searches it was assumed that the 
LSP is a neutral particle which eludes detection. This assumption leads 
to the celebrated signature of supersymmetry - missing transverse
momentum 
and energy. The recent LEP and Tevatron bounds on masses of the
supersymmetric
particles can be found in \cite{susy}{\footnote{Many of these bounds 
are based on signatures specific to certain models and do not 
\newline ${~~~~~}$ hold in general.  For example, the Tevatron bound 
on the stop mass is based on the \newline ${~~~~~}$ decay $\tilde t
\rightarrow c \tilde \chi^0$, which cannot occur in  our model.}}.

In the scenario which we are discussing, the LSP is a charged strongly
interacting particle. If it were sufficiently long-lived, it would
leave a track in the tracking chambers and thus could be detected
directly. In particular, if a heavy charged particle is produced
at the Tevatron, it can be identified in the CDF detector due
to its relatively low velocity and, consequently, large ionization
deposition, $dE/dx$. CDF measures $dE/dx$ in the silicon vertex
detector as well as in the central tracking chamber.  In Run I, no
events in excess of the expected background have been found, setting
a limit of $85\;GeV$ on the mass of a unit-charged scalar triplet
\cite{mae}.
A more recent analysis \cite{conn}
increases this bound up to about $100\;GeV$
 {\footnote{Interactions with the material of the detector
may weaken this bound by 30-40 $GeV$  \newline ${~~~~~}$ \cite{mae}.}}.

In the previous subsection we have argued that there are four metastable
shadrons. This is the case when these shadrons are in isolation.
However, when interactions with matter are turned on, only one of them
survives. Let us consider the states $\tilde t \bar u$ and
${\overline{\tilde t}} u$ propagating in matter. Apparently, $\tilde t
\bar u$ has a  larger interaction cross section since the
annihilation channel is open:
\begin{eqnarray}
\label{eq:annih}
&& \tilde t\bar u \;+\;p\;\rightarrow \tilde t ud\;
+\;\pi^0\;,\;2\pi\;,\;...\\
&& \tilde t\bar u \;+\;p\;\rightarrow \tilde t ud\;
+\;\gamma\;,\;2\gamma\;,\;...
\end{eqnarray}
As a result, the interaction eigenstates are different from the mass or
flavor eigenstates. This is a well known phenomenon in K-meson physics:
the same effect is used to regenerate the $K_S$ component via
interactions
with matter.

The effective mass matrix in the $(T^0, \bar T^0)$ basis now takes the
form
\begin{eqnarray}
&&M'= \pmatrix{m&\epsilon \cr
              \epsilon & m \cr}
\cdot \pmatrix{1-i\gamma&0 \cr
              0 & 1 \cr}
= \pmatrix{m-im\gamma&\epsilon \cr
              \epsilon -i\epsilon\gamma & m \cr}\;,
\end{eqnarray}
where we have taken into account only the largest absorptive effect.
The constant $\gamma > 0$ depends on the material and energy of the
shadron. The size of $\gamma$ can be estimated $very\; roughly$ using
the nuclear interaction length $\lambda_I$.  For protons propagating 
in iron, $\lambda_I \sim 17\;cm$ \cite{pdg}. This number triples for 
the $T^0$ as it contains only one light (anti)quark \cite{drees}, 
yielding $m\gamma \sim 10^{-15} GeV$.  Therefore, $m\gamma\ll \epsilon$ 
and the interaction eigenstates can be well approximated  perturbatively 
in $\gamma$. The corresponding effective masses are
\begin{eqnarray}
&&m_1'=(m-\epsilon)\biggl[1-{i\gamma\over2} \biggr]\;,\nonumber\\
&&m_2'=(m+\epsilon)\biggl[1-{i\gamma\over2} \biggr]\;.
\end{eqnarray}
Apparently, both mesonino states eventually vanish in a medium
converting into sbaryons. For example, in iron, the attenuation length
would be about 1 meter.  At very low energies, the mesoninos may
interact 
with matter primarily through the electromagnetic annihilation of $u$
and
$\bar u$ if the strong channel (\ref{eq:annih}) is kinematically
prohibited.

Similar considerations can be applied to the sbaryon $S^-$ which
contains light antiquarks. As a result, the only stop-containing
particle which does not get destroyed in matter is $S^+$. All other 
shadrons propagating in a medium will go through a series of reactions 
and eventually end up as $S^+$.

Nevertheless, if the $\lambda'$ and $\lambda''$ are anywhere
within 10 orders of
magnitude from the lower bound (\ref{eq:bound}), all of the charged
shadrons but $V$ (which decays too quickly) would leave tracks in the
tracking chambers. Since they are expected to be very massive, they
would pass through the electromagnetic calorimeter undetected.
With some energy loss (which may be considerable for slow particles
\cite{drees}) in the hadron calorimeter, a shadron
will penetrate the detector and will be triggered on as a muon.

Let us consider the interaction of shadrons with the hadron
calorimeter in more detail.  In collisions with the nucleons, the 
momentum transfer at typical Tevatron energies is very small and 
the energy available in the center of mass system is likely to be 
below the pion threshold \cite{mae}. Thus, the inelastic scattering
would primarily go through the annihilation of the light quarks and 
antiquarks as in Eq.\ref{eq:annih}. Since the $S^+ - T^0$ mass
difference 
is estimated to be around $400\;MeV$, a few soft pions (or an equivalent
amount of radiation) can be produced in the collision of $T_{1,2}$ and
$p$.
Note that, at low energies, only the shadron states involving light 
antiquarks ($T^+$,$\;S^-$,$\;etc.$) would deposit  any tangible amount
of energy in the hadron calorimeter. However, at the Tevatron, the
deposited energy is unlikely to exceed a few $GeV$ in any case and thus
the signal would be hard to distinguish from that generated by a charged 
colorless LSP such as a stau, especially taking into account that the
energy
resolution is only $80-100\% / \sqrt{E}$ \cite{carena}.

The situation will improve at the LHC. The energy deposition in the
calorimeter will be sufficient to distinguish between a strongly and 
a weakly interacting LSP. The  energy loss in a material as a function 
of the mass and energy of a strongly interacting particle was studied in
Ref.\cite{moh} based on kinematical considerations. For instance, it was
shown that a $100\;GeV$ particle at $E=1\;TeV$ would trigger a few 10
$GeV$ hadron showers. Also, it was noted that, due to a slower velocity,
the opening angle of the shower would be significantly wider than that 
of the shower induced by regular hadrons. These calorimeter data
combined 
with the tracking (and timing) information would allow to eliminate
the SM background and determine the nature of a new (meta)stable
particle.

Let us now briefly discuss other implications of the scenario.  The 
stop may be sufficiently long-lived as to be treated as stable.  One may 
ask whether there could be $non$-$collider$ experiments which can detect
shadrons produced in the collisions.  In principle, this can be done via
mass spectroscopy. If a produced shadron is sufficiently slow, it loses 
most of its energy through ionization and thus can be trapped in the
metal
plates of the detector \cite{drees}, eventually ending up as an $S^+$.
The isosinglet $S^+$ is very unlikely to be bound in nuclei due to the
Coulomb repulsion and absence of the one-pion-exchange potential. 
Therefore, 
it should be found in isolation. Having picked up an electron, it will
form 
an anomalously heavy hydrogen atom. If such atoms existed in an
appreciable 
concentration, they would be detectable by mass spectrometry
\cite{smith} 
as they possess an anomalously small charge/mass ratio.

We would like to conclude with a remark:
\begin{center}
``Curiosity is a delicate little plant which, aside from \\
stimulation, stands mainly in need of freedom.''  \\
$~~~~~~~~~~~~~~~~~~~~~~~~~~~~~~~~~~~$    ---- Albert Einstein
\end{center}

We have benefited from very helpful correspondence with T. Cohen, H.
Dreiner,
X. Ji, A. Kronfeld, D. Stuart as well as from discussions with members
of
Virginia Tech and U. of Maryland high energy groups. We are particularly
thankful to  Lay N. Chang, S. Eno, R.N.Mohapatra and G. Snow for
enlightening
discussions.

\newpage

\input FEYNMAN
\vskip 1.5in
\begin{picture} (18000,8000)
\THICKLINES
\drawline\fermion[\E\REG](11000,0)[13000]
\put(16000,800){$\tilde g\;,\;\tilde \chi^0$}
\THINLINES
\drawline\fermion[\NW\REG](\fermionfrontx,\fermionfronty)[5948]
\drawarrow[\LDIR\ATTIP](8026,2974)
\put(5526,\fermionbacky){$u$}
\drawline\scalar[\SW\REG](\fermionfrontx,\fermionfronty)[3]
\put(5526,\scalarbacky){$\tilde t$}
\drawline\fermion[\NE\REG](24000,0)[5948]
\drawarrow[\LDIR\ATTIP](8026,-2974)
\drawarrow[\LDIR\ATTIP](26974,2974)
\put(28974,\fermionbacky){$u$}
\drawline\scalar[\SE\REG](24000,0)[3]
\drawarrow[\NW\ATTIP](26974,-2974)
\put(28974,\scalarbacky){$\tilde t$}
\end{picture}
\vskip 0.9in
\begin{center}
{\rm Figure 1: $T^0-\bar T^0$ mixing.}
\end{center}

\vskip 0.8in
\begin{picture} (18000,8000)
\THICKLINES
\drawline\fermion[\E\REG](11000,0)[13000]
\put(17500,-300){${\bf \times}$}
\put(17500,1000){$m_{\tilde g}$}
\THINLINES
\drawline\fermion[\NW\REG](\fermionfrontx,\fermionfronty)[5948]
\drawarrow[\LDIR\ATTIP](8026,2974)
\put(5226,\fermionbacky){$u_R$}
\drawline\scalar[\SW\REG](\fermionfrontx,\fermionfronty)[3]
\put(5226,\scalarbacky){$\tilde t_R$}
\drawline\fermion[\NE\REG](24000,0)[5948]
\drawarrow[\LDIR\ATTIP](8026,-2974)
\drawarrow[\LDIR\ATTIP](26974,2974)
\put(28974,\fermionbacky){$u_R$}
\drawline\scalar[\SE\REG](24000,0)[3]
\drawarrow[\NW\ATTIP](26974,-2974)
\put(28974,\scalarbacky){$\tilde t_R$}
\end{picture}
\vskip 0.9in
\begin{center}
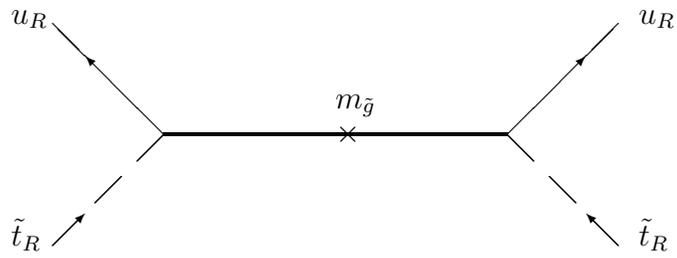
{\rm Figure 2: The dominant contribution to $T^0-\bar T^0$ mixing.}
\end{center}

\end{document}